# A Preliminary Study on Aging Examining Online Handwriting

Marcos Faundez-Zanuy∗, Enric Sesa-Nogueras∗, Josep Roure-Alcobé∗, Anna Esposito†,
Jiri Mekyska‡ and Karmele López-de-Ipiña§
∗Tecnocampus
Avda. Ernest Lluch 32
08302 Matar (Barcelona), Spain
Email: {faundez, sesa, roure }@tecnocampus.cat
†Seconda Universita` di Napoli and IIASS, Italy
Email: iiass.annaesp@tin.it
‡Brno University of Technology, Brno, Czech Republic
Email: j.mekyska@phd.feec.vutbr.cz
§Basque Country University, Donostia, Spain
Email: karmele.ipina@ehu.es

*Abstract*—In order to develop infocommunications devices so that the capabilities of the human brain may interact with the capabilities of any artificially cognitive system a deeper knowledge of aging is necessary. Especially if society does not want to exclude elder people and wants to develop automatic systems able to help and improve the quality of life of this group of population (healthy individuals as well as those with cognitive decline or other pathologies).

This paper tries to establish the variations in handwriting tasks with the goal to obtain a better knowledge about aging. We present the correlation results between several parameters extracted from online handwriting and the age of the writers. It is based on BIOSECURID database, which consists of 400 people that provided several biometric traits, including online handwriting. The main idea is to identify those parameters that are more stable and those more age dependent. One challenging topic for disease diagnose is the differentiation between healthy and pathological aging. For this purpose, it is necessary to be aware of handwriting parameters that are, in general, not affected by aging and those who experiment changes (increase or decrease their values) because of it. This paper contributes to this research line analyzing a selected set of online handwriting parameters provided by a healthy group of population aged from 18 – 70 years. Preliminary results show that these parameters are not affected by aging and therefore, changes in their values can only be attributed to motor or cognitive disorders.

## I. INTRODUCTION

A challenging research topic is the differentiation between healthy and pathological individuals, considering that health is something that evolves with aging. While most interesting research is based on longitudinal studies, this is time consuming and difficult to implement, due to the impossibility to ensure the availability of participants during a long time period.

Aging can be defined as the accumulation of changes in people over time [1]. Aging is a multidimensional process of physical, psychological, and social change. Some dimensions of aging grow and expand over time, while others decline. Reaction time, for example, may slow with age, while knowledge of world events and wisdom may expand. Research shows that even late in life, potential exists for physical, mental, and social growth and development [2]. Aging is an important part of all human beings reflecting the biological changes that occur, but also reflecting cultural and societal conventions. Aging is among the largest known risk factors for most human diseases [3]. Roughly 100.000 people worldwide die each day of age-related causes [4].

While aging affects daily life activities, it also affects the interaction capabilities with other people as well as machines.

Cognitive infocommunications (CogInfoCom) [5]–[7] investigates the link between the research areas of infocommunications and the cognitive sciences, as well as the various engineering applications which have emerged as the synergic combination of these sciences.

The primary goal of CogInfoCom is to provide a systematic view of how cognitive processes can co-evolve with infocommunications devices so that the capabilities of the human brain may not only be extended through these devices, irrespective of geographical distance, but may also interact with the capabilities of any artificially cognitive system. This merging and extension of cognitive capabilities is targeted towards engineering applications in which artificial and/or natural cognitive systems are enabled to work together more effectively.

From this point of view some interesting possibilities appear, such as the improvement of those human beings experiencing some cognitive decline with the goal to provide a successful aging. This requires some measurement functions, which could be based on high level activities done by people (speech, handwriting, etc.).

The concept of successful aging can be traced back to the 1950s and was popularized in the 1980s. Successful ageing consists of three components [8]:

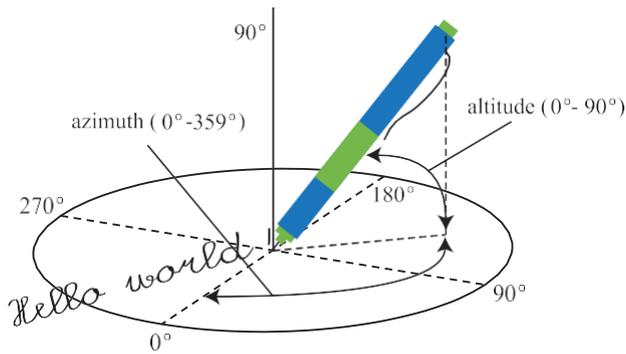

Fig. 1. Azimuth and inclination angles of the pen with respect to the plane of the graphic card

1) Low probability of disease or disability
2) High cognitive and physical function capacity
3) Active engagement with life.

In this paper we will use a simple approach, consisting of evaluating a selected set of online handwriting parameters extracted from the benchmarked handwriting database, BIOSECURID [9]. The database contains the handwriting of 400 subjects, which were required to perform specific handwriting tasks (among those their signature and copying a predefined paragraph). The handwritten was acquired by means of an Intuos Wacom 4 digitizing tablet plus an inkpen. From the users point of view, he is writing with a ordinary ball pen on an ordinary sheet of paper. In fact, the digitizing tablet is behind the sheet. The nice advantage of this online handwriting acquisition is the possibility to accurately measure handwriting timing, pressures and angles. In addition, the handwriting data are acquired in real time.

Other than the Intuos Wacom digitizing tablet, online handwriting data can be acquired through a stylus-operated PDAs. These devices can capture the following information:

1) Position in x-axis.
2) Position in y-axis.
3) Pressure applied by the pen.
4) Azimuth angle of the pen with respect to the tablet (see Fig. 1).
5) Altitude angle of the pen with respect to the tablet (see Fig. 1).

Using this set of dynamic data, further information can be inferred, such as handwriting acceleration, velocity, instantaneous trajectory angle, instantaneous displacement, tangential acceleration, curvature radius, centripetal acceleration, and more.

In order to establish a baseline for comparing healthy vs. pathological handwriting parameters (because of aging) it is necessary to have information on how they change along the time. To this aim, the authors collected a multimodal database of healthy people, which consists of handwriting data provided by university students, lecturers and administrative/support people. On the other hand, the acquisition of pathological samples requires:

- The participation of medical doctors able to label the samples.

Fig. 2. Cursive handwriting task produced by a male subject

- The access to people affected by some pathology.

For most of the engineering teams it is hard to access this kind of samples, mostly because of ethical and privacy issues. For these reasons, to date, online pathological handwriting databases do not exist. Nevertheless, the differentiation between pathological and healthy samples requires a previous analysis of healthy population. This paper wants to contribute to this last issue.

## II. EXPERIMENTAL RESULTS

### A. Biosecurid Database

In this paper we use a specific task of the BIOSECURID database [1], which consists of the copy of a predefined paragraph. Fig. 2 shows an example obtained from a masculine writer.

Using this task we have evaluated the parameters described in table I for the 400 users and for second acquisition session (the database consists of four different acquisition sessions suitable to analyze intra-user variations, which are not considered in this study). For sake of clarity, it must be said that the majority of the subjects involved in the data collection were university students with an average age of 20 years, as it can be seen from Fig. 3 that illustrate the subjects' age histogram. Nonetheless, the database also contains the handwriting of more aged subjects and their data were used for comparison.

### B. Experimental Setup

A Pearson correlation between several features extracted from the online handwritten task and the age of the writer is applied, to check if some correlation exists.

In statistics, the Pearson product-moment correlation coefficient (sometimes referred to as the PPMCC or PCC or Pearson's r) is a measure of the linear correlation (dependence) between two variables X and Y, giving a value between +1 and −1 inclusive, where 1 is total positive correlation, 0 is no correlation, and −1 is total negative correlation. It is widely used in sciences as a measure of the degree of linear dependence between two variables. In our case, the first variable is a measurement performed on the handwriting task (for instance time required on the paper surface to finish the task) and the second variable is the age of the writer. The interpretation of the correlation coefficient can be based on these ranges:

- High correlation: .5 to 1.0 or −0.5 to 1.0
- Medium correlation: .3 to .5 or −0.3 to .5
- Low correlation: .1 to .3 or −0.1 to −0.3

C. Experimental Results

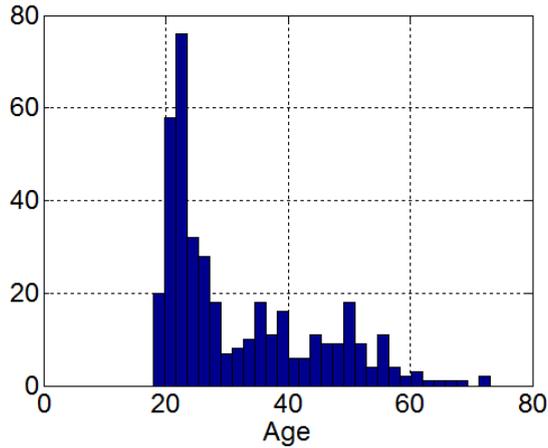

Fig. 3. Histogram of the participants' ages

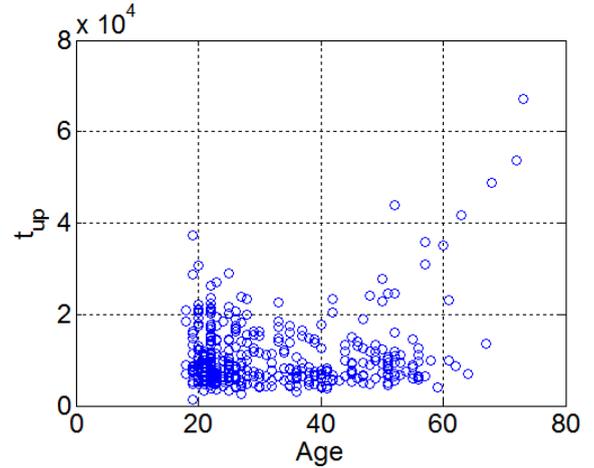

Fig. 4. Distribution of the pair of values (age, time up) for each writer

Fig. 4 illustrates the distribution along the age of the handwriting parameter "time up", i.e. the time the pen was not on the sheet of paper (age, time up), for each writer. Fig. 5 represents the same for the "times down" parameter.

Table I shows the selected set of handwriting measured features, their Pearson correlation coefficinet and the p value, which is an index of the correlation significance. The selected handwriting features are: time up and down, pressure, speed of the trajectory, entropy, Zero crossing rate, number of strokes, normalized times, differential values obtained by the first (d) and second derivative (dd), pressure higher than a predefined threshold and Teager energy operators [10]. The *m* at the end of the parameter name stands for mean, *std* for standard deviation, and also reported are the median, mode, mean value (m) etc.

As it can be read from the "Rho" column reported in Table I, none of the selected handwriting parameters show a high correlation with the age, suggesting that age do not affect this daily functional activity. For those parameters where it would be possible to guess a weak correlation, the *p* value was too high to made it significant. A weak correlation with the age (near medium correlation) was found only for the "nt_up" feature (see Table I for the list of selected parameters), which has been defined by the authors as the normalized time up. It consists of the time up split by the number of strokes up in the air.

While our initial guess was that some of the selected features could have been affected by aging, the results of this research do not confirm it. However, these results have very good implications: once it has been established that the selected handwriting parameters are not (or weakly) affected by aging, they can be used as a baseline for discriminating between healthy/pathological subjects. Instead, if our initial guess was confirmed, the healthy/pathological classification of subjects through handwriting parameter would have be more challenging, suggesting that handwriting features were not useful for discriminating among healthy and pathological subjects.

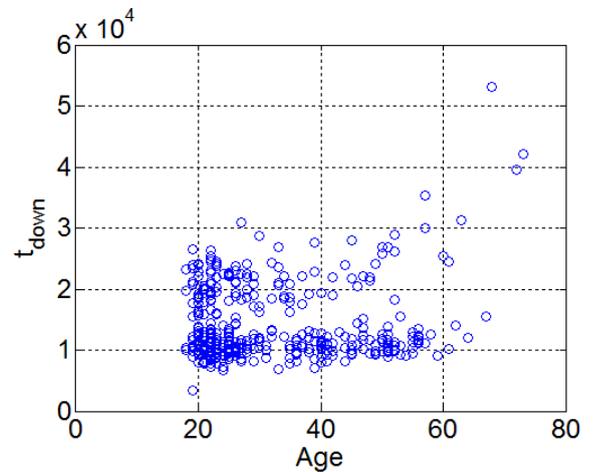

Fig. 5. Distribution of the pair (age, time down) for each writer

Fig. 4 and 5 show some values in the region of the "more aged" area that seem to suggest an increase of time required to do the task, especially in the "*up in the air*" case. However, as it can be seen from the data reported in Table I, this is not confirmed by a strong Person's correlation coefficient. Clearly, these results must be confirmed with more data, in particular with more data from more aged subjects. Thus, we should consider these results preliminary.

To substantiate these results, one future work is to acquire a balanced database containing handwriting data from a large and balanced range of ages, and from individuals with similar education level, and maybe also from different culture.

TABLE I. SOME EXTRACTED FEATURES, THEIR PEARSON'S CORRELATION AND THE *p* VALUE.

| features | Rho | p |
|---|---|---|
| t_upm | 0.2 | 6.80E-05 |
| t_downm | 0.14 | 0.00448107 |
| p_meanm | 0.1 | 0.03966636 |
| p_maxm | 0.12 | 0.02060574 |
| p_medianm | 0.09 | 0.05866625 |
| p_modem | 0.08 | 0.12951908 |
| p_stdm | 0.06 | 0.22729248 |
| speed_maxm | 0.06 | 0.2198899 |
| entropy_xm | -0.09 | 0.05843317 |
| entropy_ym | 0.12 | 0.01604943 |
| entropy_pm | -0.1 | 0.05707732 |
| ZCRm | -0.21 | 2.73E-05 |
| NZCRm | -0.21 | 2.73E-05 |
| strokes_dm | -0.21 | 2.73E-05 |
| strokes_um | -0.21 | 2.73E-05 |
| nt_up | 0.29 | 4.50E-09 |
| nt_down | 0.22 | 5.86E-06 |
| dp_meanm | 0.04 | 0.404588 |
| dp_maxm | 0.19 | 0.00016188 |
| ddp_maxm | 0.18 | 0.00036086 |
| entropy_dpm | -0.19 | 0.00016748 |
| entropy_ddpm | -0.19 | 9.04E-05 |
| entropy_accelerationm | -0.06 | 0.23452631 |
| p100m | 0.16 | 0.00157878 |
| p200m | 0.17 | 0.00069852 |
| p300m | 0.15 | 0.00188938 |
| p400m | 0.12 | 0.01359625 |
| p500m | 0.12 | 0.01319046 |
| p600m | 0.14 | 0.00625067 |
| p700m | 0.14 | 0.00471245 |
| p800m | 0.13 | 0.00774605 |
| p900m | 0.12 | 0.02098432 |
| teagerxmax | 0.04 | 0.42055342 |
| teagerym | 0.1 | 0.04343131 |
| teagerymedian | -0.02 | 0.7450191 |
| teagerymax | 0.11 | 0.02860634 |
| teagerpm | 0.01 | 0.87595643 |
| teagerpmedian | -0.01 | 0.90101579 |
| teagerpmax | 0.15 | 0.00216337 |

III. CONCLUSION

In this paper we have analyzed online handwriting features to search for age dependence. However, the Pearson correlation values for the parameters under examination revealed to be quite low, and not significant. While these results discourage the existence of a specific handwriting parameter indicative of age, we provided with these results a baseline for discriminating between healthy and pathological online handwriting parameters, considering that now we can hypothesize that handwriting is only affected by motor and cognitive disorders.

In addition, we should take into account that biological age does not need to be related to the "apparent age". Some people seem younger/older than they really are.

These findings also imply that the proposal of an automatic age estimator based on handwritten tasks is not a trivial problem, but it also implies that the alteration of some of these parameters could be related to health issues regardless of the age of the writer.

Future research works could be devoted to investigate the effects of age on other biometric traits also available in BIOSECURID, such as speech and face changes.

ACKNOWLEDGMENT

This work has been supported by FEDER and Ministerio de ciencia e Innovacin, TEC2012-38630-C04-03, COST IC 1206, GAP102/12/1104 and NT13499. The described research was performed in laboratories supported by the SIX project; the registration number CZ.1.05/2.1.00/03.0072, the operational program Research and Development for Innovation.